# Observation of Wigner cusps in a metallic carbon nanotube

Brandon Blue[1,2], Ryuichi Tsuchikawa[1,2], Amin Ahmadi[1], Zhengyi Zhang[4], Daniel Heligman[1,2], Stephanie D. Lough[1,2], James Hone[4], Eduardo R. Mucciolo[1], Masa Ishigami[1,2,*]

1. Department of Physics, University of Central Florida, Orlando, FL 32816
2. NanoScience Technology Center, University of Central Florida, Orlando, FL 32816
3. Department of Mechanical Engineering, Columbia University, New York, NY 10027

*Corresponding author: ishigami@ucf.edu, Department of Physics, University of Central Florida, 4111 Libra Drive PS 430, Orlando, FL 32816, USA

Abstract

Previous gate-dependent conductance measurements of metallic carbon nanotubes have revealed unexplainable conductance suppressions, occurring at two different gate voltages. These were previously attributed to the gate-dependency of contact resistance. Our gate-dependent conductivity measurements on a metallic nanotube with known chirality show that these bimodal conductance suppressions are the manifestations of Wigner cusps, often seen in atomic and nuclear physics experiments.

## 1. Introduction

Wigner cusps manifest in nuclear and atomic physics in processes such as photodetachment, photoionization, and inelastic atomic collisions [1, 2]. Wigner cusps are enhancements in the scattering cross section at the threshold energies where new scattering channels are opened [3]. Wigner cusps are hardly ever discussed in condensed matter physics. They have only been discussed in a few theoretical studies [4] and have not been clearly identified in experiments although suggestive features have been reported [5, 6].

We determine gate-dependent conductivity of a long, isolated metallic carbon nanotube with known atomic structure (i.e. chirality) using the transfer length method [7-9]. Specifically, the conductivity is determined by using four devices with varying channel lengths fabricated on the same nanotube as a function of gate voltage. We find the conductivity to be suppressed at multiple gate bias voltages, unexpected for metallic carbon nanotubes. We perform in-depth analysis of the gate-dependent conductivity in comparison to our theoretical calculations using the recursive Green's function method within the Landauer formalism [10]. We are able to conclude that the scattering induced by extrinsic charged impurities, specifically charge traps in device substrates, is enhanced when the second subband becomes available for carrier transport. As such, our observation of the unexpected gate-dependent conductivity suppressions represent the manifestation of Wigner cusps in a metallic carbon nanotube.

## 2. Material and Method

This paper focuses on a representative metallic carbon nanotube with a chiral index of (22,4). A single-wall carbon nanotube is grown suspended over a 60-μm trench by chemical vapor deposition [11]. Rayleigh scattering spectroscopy is performed to identify its chirality [12]. Once the chirality is identified, the nanotube is transferred to a $SiO_2$/Si substrate with 280 nm thermal oxide. We used the same device fabrication procedure as described in our previous work [9]. Gold contacts are used to minimize contact resistance. Multiple contacts are fabricated on the nanotube, resulting in transport channels of different lengths, to enable the transfer length method [7-9] and, therefore, the determination of conductivity. Resulting devices are annealed in a flow of $Ar/H_2$ at 360 °C for 3 hours to remove polymer residues and other contaminants from the fabrication process [13]. Finally, the devices are annealed in ultra high vacuum at 200 °C for 12 hours prior to measurements performed at room temperature.

## 3. Results and Discussion

Figure 1a shows gate dependent conductance of a 2 μm long segment of the (22,4) metallic carbon nanotube. Gate voltage is first ramped down to the most negative set point. Measurements are performed while ramping up gate voltage to the most positive set point, then while ramping down to the most negative voltage. The device manifests an unusual gate dependence, unexpected for metallic carbon nanotubes. These unusual gate dependencies have also seen by others, but have been dismissed as being due to contact resistance [14]. A large hysteresis is also observed. Changes in the charge environment with negative charge being injected into silicon oxide at positive gate biases and positive charge being injected into the dielectric at negative gate biases are known to induce such hysteresis [15, 16]. Figure 1b shows upward gate-dependent conductances of 4 devices on the *same* (22,4) metallic nanotube with different device lengths as a function of gate voltage. Conductance of all segments are minimized near $V_{gate}$ = -20 V, and again suppressed above $V_{gate}$ = -5 V. The conductance remains suppressed above $V_{gate}$ = -5V, and the amount of suppression becomes larger for longer segments. The observed bimodal conductance suppressions cannot be explained by intrinsic mechanisms in metallic carbon nanotubes such as the curvature-induced bandgap [17] or the Mott insulating state [18], which both suppress the conductance at only one gate voltage, corresponding to the charge neutrality point. As such, previous observations of similar bimodal conductance suppressions have been attributed to the dependency of contact resistance on gate voltage [14].

Figure 2 shows the gate dependent conductivity of the same nanotube determined by using the transfer length method [7-9], in which resistivities are determined at each gate voltage by fitting measured resistances as a function of lengths. The conductivity is minimized near $V_{gate}$ = 10 and 30 V in the downward scan shown in Figure 2a while the upward scan, shown in Figure 2b, has minimum near -20 V and conductivity suppression starting at 0 volt and continuing to higher voltages. The presence of the bimodal suppression in the *conductivity* data shows that it is not due to contact resistance. Interestingly, the observed gate dependencies for downward and upward scans are not simply shifted in gate voltage. The bimodal suppression feature is much broader for the upward scan and the downward scan appears approximately symmetric across the maximum conductivity. This indicates that the sign of charge of traps in the substrate sensitively affects the bimodal conductivity suppression since gating modifies the injected charge [15, 16].

We calculate the energy dependence of the conductance using the recursive Green's function method within the Landauer formalism in order to understand the origin of the bimodal conductivity suppression. We simulate diffusive transport in a (22,4) metallic carbon nanotube by randomly placing scatterers on the nanotube in the form of local Gaussian potential fluctuations $V(r) = \sum_i V_0 \exp\left[-\frac{(r-r_i)^2}{2\xi^2}\right]$, where $r_i$ is the scatterer random location. 200 distinct numerical samples were created in this fashion. The conductance for each sample is computed and averaged to suppress quantum fluctuations. The chiral index determines the energies at which the second subband onsets and our analysis indicates that the second subband onsets at ± 0.582 eV with respect to the charge neutrality point for a (22,4) nanotube. Densities of states calculations [19] with the capacitance of $SiO_2$ dielectric used for the experiments indicates that the onset for the second subband will occur at ± 10.0 volts with respect to the charge neutrality point.

Figure 3 shows the conductance of a nanotube with a length of 1.78 μm as a function of energy with long-ranged scatterers, which possess $V_0$ = 1.4 eV and ξ = 10 Å, to simulate charge traps in silicon oxide [20-23]. Our focus on the long-ranged scatterers is based on the observed behavior in the gate hysteresis measurements, which indicate that charge traps play a pivotal role in producing the observed anomalous conductivity suppressions. Furthermore, they are the most likely scatterers present in our experiment carried out on a clean nanotube in ultra high vacuum. Our theoretical calculations coincide the charge neutrality point at the zero gate voltage, which may be different from the experiment where such coincidence cannot be guaranteed. We chose the number of randomly distributed impurities to be 13, which corresponds to the impurity density of $3.8 \times 10^{11}/cm^2$, the same order of magnitude as the density of charge traps in silicon oxide [20-23]. Since negative charge is injected into silicon oxide at positive gate biases and positive charge is injected into the dielectric at negative gate biases [15, 16], applying high gate bias voltage can effectively change the net charge of the trapped charge, which can act as scatterers. As such a result, ramping up the gate bias from high negative value will induce the net charge of the charge traps in silicon oxide to be more positive than ramping down the gate bias from high positive value.

As such, we show our calculated results for positively charged scatterers, which simulates the case when the net charge of charge traps is positive, as well as for equal numbers of positive and negative scatterers, which simulates the case for when the net charge of charge traps is neutral. We note that the negatively charged scatterers produce the gate-dependent conductance that is mirror reflected across the charge neutrality point when compared to the case of the positively charged scatterers.

Our theoretical calculations show bimodal conductance suppression in the presence of net neutral scatterers as shown in Figure 3a. The gate-dependent conductance is found to be mirror symmetric across the charge neutrality point located at $V_{gate} = 0$ V for our theoretical calculations. When only the first subband of the metallic nanotube participates, with $|V_{gate}| < 10$ V, pseudospin conservation in metallic carbon nanotubes [24] prevents most scattering as expected. Introduction of the second subband is expected to double the conductance to 8 $e^2/h$ without scattering. However, the conductance is suppressed near $V_{gate} = \pm 10$ V with the minima located precisely at the onset of the second subbands. The conductance recovers slightly at higher $|V_{gate}|$, indicating that the net neutral scatterers have the strongest impact precisely when the new scattering channels are imparted by the access to the second subbands, just as predicted for Wigner cusps. As such, we conclude that the conductance suppression at $V_{gate} = \pm 10$ V is the manifestation of Wigner cusps. The conductance remains well below 8 $e^2/h$ in the second subband because of intersubband scattering. Figure 3b shows our calculations for positive scatterers. A Wigner cusp is again observed at $V_{gate} = -10$ V, but the conductance suppression is much weaker at $V_{gate} = +10$ V. This result indicates that scattering of electrons by positive scatterers is insufficient to induce a Wigner cusp. Such result is expected because positive scatterers induce potential barriers for holes and potential wells for electrons and, as observed previously by us [9], holes are significantly more susceptible to scattering by positive scatterers. As a result, the conductance is only suppressed far above $V_{gate} = +10$ V when the intersubband scattering starts to contribute more significantly, forming the bimodal suppression feature with a much broader profile than neutral scatterers.

These calculated gate dependencies match our experimental data. The impact of net neutral scatterers in numerical simulations is consistent to the measured downward conductivity with which the conductance minimums are separated by ~20 volts and the gate dependency between the conductance minimums is symmetric across the maximum conductance point as shown in

Figure 2a. If the calculated result for net neutral scatterers corresponds to the downward conductivity, the case for positive scatterers is expected to match to the measured upward conductivity. Indeed, the measured upward gate dependency is observed to be wider and non-symmetric across the maximum conductance point as shown in Figure 2b in consistency with the calculated dependency for positive scatterers. These comparisons between the theory and the experiment on the gate dependent conductivity and on the hysteresis lead to the conclusion that the observed bimodal conductivity suppressions in metallic nanotubes are due to the manifestation of Wigner cusps.

## 4. Conclusion

We have shown that the charged impurities can induce a bimodal conductance suppression in metallic carbon nanotubes. Our measurement and theoretical calculations show that these conductance suppressions are Wigner cusps. Our observation represents one of the few observations of Wigner cusps in condensed matter physics.

## Acknowledgements

This work was supported by the National Science Foundation under the Grants No. 1006230 and No. 1006533.

**Figure Captions**

Figure 1 (a) Conductance measured for a 2 μm segment of the (22,4) nanotube. (b) Conductance measured for 2, 3, 6, and 10 μm segments of the (22,4) nanotube with the gate voltage scanned upward from -30 to 30 V.

Figure 2 Conductivity of the (22,4) nanotube determined from the conductances for 2, 3, 6, and 10 μm segments with the gate voltage swept (a) downward and (b) upward.

Figure 3 Calculated conductance for 1.78 μm segment of a (22,4) nanotube with 13 impurities. Results for (a) net neutral scatterers and for (b) positive scatterers are shown.

Figure 1

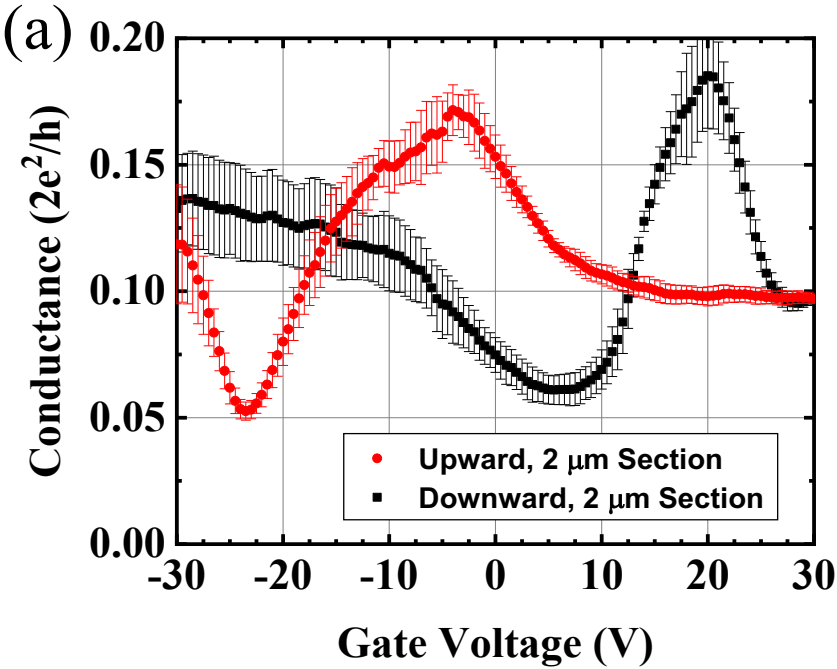 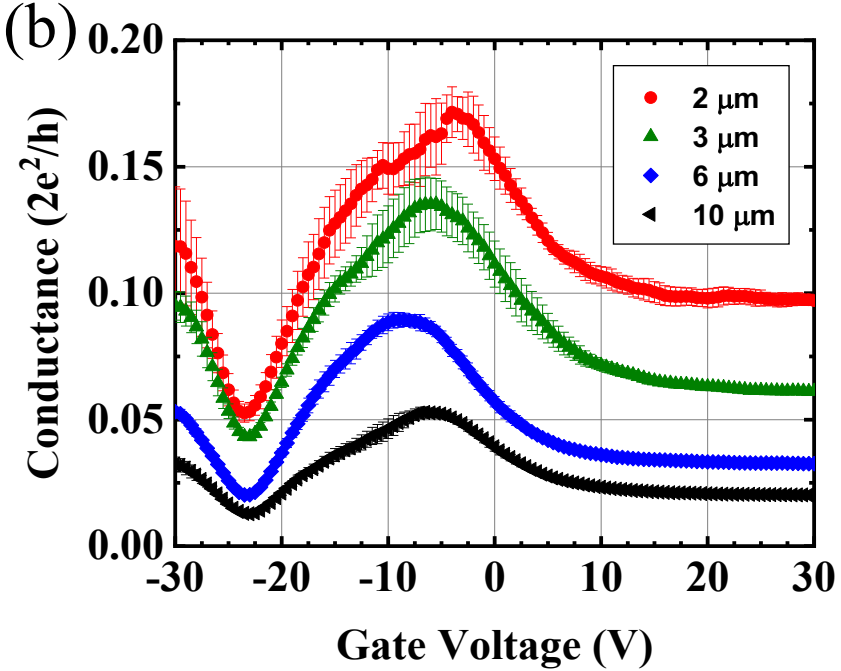

Figure 2

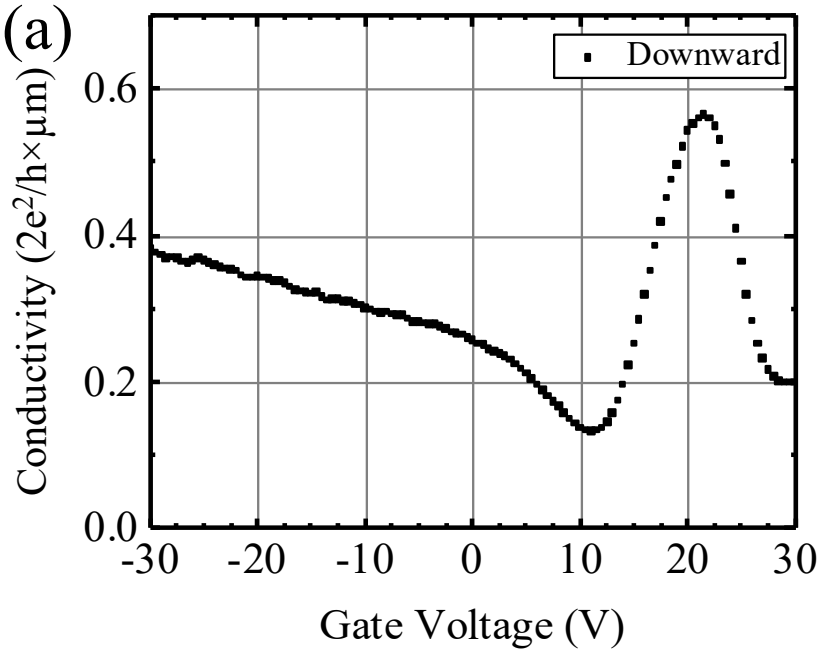 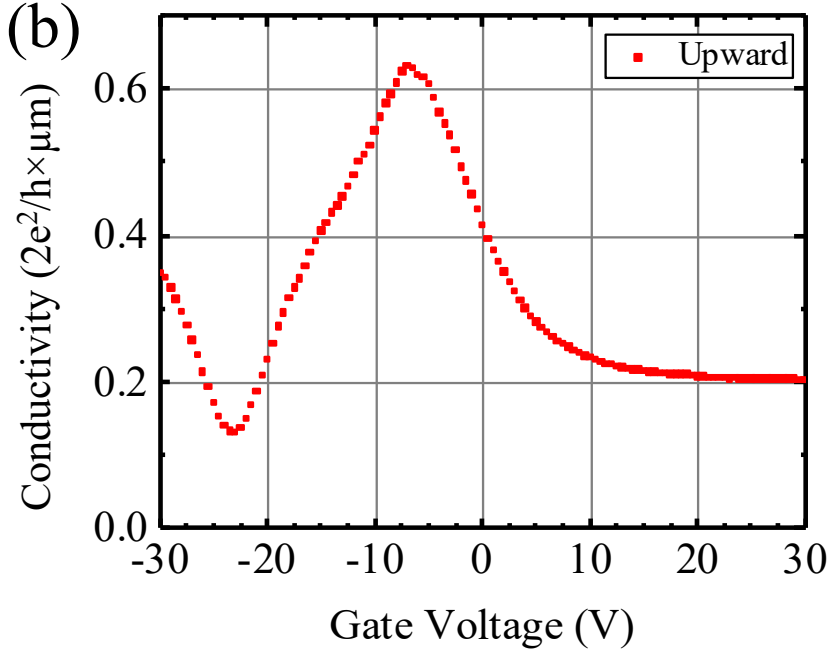

Figure 3

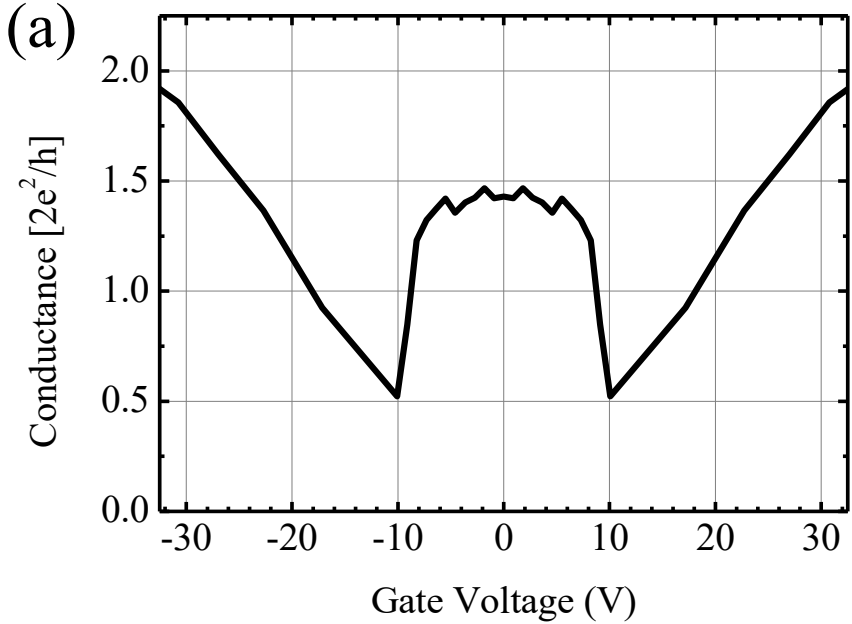 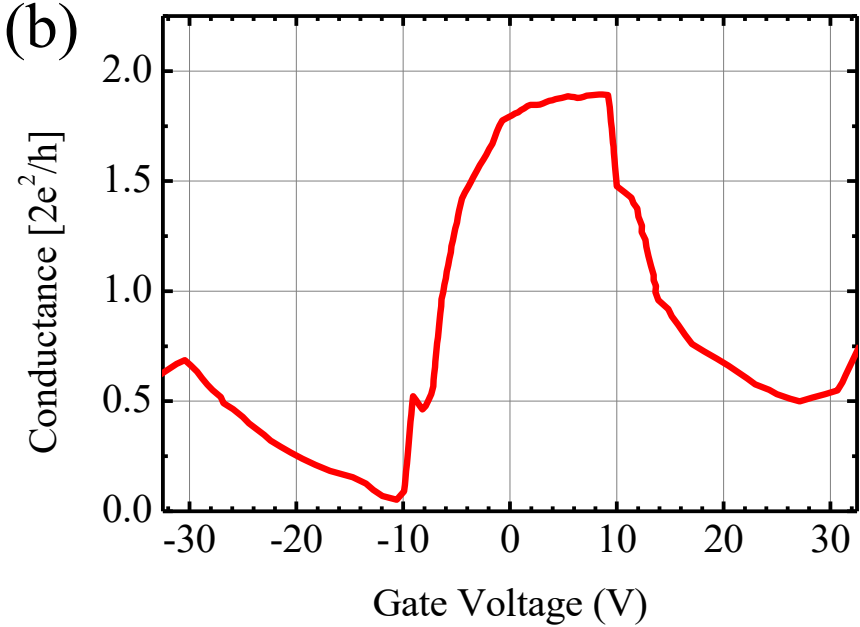